\documentclass[mnsc,noblindrev]{informs3}
\OneAndAHalfSpacedXI

\usepackage{natbib}
 \bibpunct[, ]{(}{)}{,}{a}{}{,}%
 \def\bibfont{\small}%
\TheoremsNumberedThrough     
\ECRepeatTheorems

\EquationsNumberedThrough    

\usepackage{amssymb}
\usepackage{graphicx}
\usepackage{amsmath}
\usepackage{algorithm}
\usepackage[noend]{algpseudocode}
\usepackage{subfig}
\usepackage{url}
\usepackage{color}
\usepackage{multirow}
\usepackage{babel}
\usepackage[shortlabels]{enumitem}
\usepackage{url}
\urldef{\mailsa}\path|{alfred.hofmann, ursula.barth, ingrid.haas, frank.holzwarth,|
\urldef{\mailsb}\path|anna.kramer, leonie.kunz, christine.reiss, nicole.sator,|
\urldef{\mailsc}\path|erika.siebert-cole, peter.strasser, lncs}@springer.com|

\DeclareMathOperator{\EX}{\mathbb{E}}

\renewcommand{\theARTICLETOP}{}
\usepackage[left=1.0in,right=1.0in,top=1.0in,bottom=1.0in]{geometry}
\usepackage{fancyhdr} 
\makeatletter
\let\c@table\c@figure %
\let\ftype@table\ftype@figure %
\makeatother

\fancypagestyle{firstpage}
{
    \fancyhead[L]{Journal of Financial Data Science, 3 (1): 10-27}    
    \fancyhead[R]{}
}

\pdfoutput=1

\usepackage{calc}
\begin{document}

\pagestyle{plain} 

\RUNTITLE{Deep Hedging of Derivatives Using Reinforcement Learning}

\TITLE{Deep Hedging of Derivatives Using Reinforcement Learning}

\ARTICLEAUTHORS{%
\AUTHOR{Jay Cao, Jacky Chen, John Hull, Zissis Poulos\thanks{We thank Ryan Ferguson, Ivan Sergienko, and Jun Yuan for helpful comments. We also thank the Rotman Financial Innovation Lab (FinHub) and the Global Risk Institute in Financial Services for support}}
\centerline{Joseph L. Rotman School of Management, University of Toronto}
\centerline{\{jay.cao, jacky.chen17, john.hull, zissis.poulos\}@rotman.utoronto.ca
}
\centerline{December 2019. This Version: September 2020}}

\ABSTRACT{

\noindent This paper shows how reinforcement learning can be used to derive optimal hedging strategies for derivatives when there are transaction costs. The paper illustrates the approach by showing the difference between using delta hedging and optimal hedging for a short position in a call option when the objective is to minimize a function equal to the mean hedging cost plus a constant times the standard deviation of the hedging cost.  Two situations are considered. In the first, the asset price follows a geometric Brownian motion.  In the second, the asset price follows a stochastic volatility process. The paper extends the basic reinforcement learning approach in a number of ways. First, it uses two different Q-functions so that  both the expected value of the cost and the expected value of the square of the cost are tracked for different state/action combinations. This approach increases the range of objective functions that can be used. Second, it uses a learning algorithm that allows for continuous state and action space. Third, it compares the accounting P\&L approach (where the hedged position is valued at each step) and the cash flow approach (where cash inflows and outflows are used). We find that a hybrid approach involving the use of an accounting P\&L approach that incorporates a relatively simple valuation model works well. The valuation model does not have to correspond to the process assumed for the underlying asset price.
}%

\maketitle

\section{Introduction} \label{intro}

Hedging is an important activity for derivatives traders. Suppose a trader sells a one-year European call option on 10,000 shares of a non-dividend-paying stock when the stock price is \$100 and the strike price is \$100. If the volatility of the stock is 20\%, the price of the option, assuming that the stock price follows geometric Brownian motion and the risk-free interest rate is 2\%, is about \$90,000. However, in the absence of hedging the trader is exposed to risk if the option is sold for \$90,000.  A two-standard-deviation upward move in the stock price during the year would cost the trader much more than the price charged.

Traders have traditionally hedged the risks associated with derivatives transactions by monitoring  ``Greek letters.'' Delta, the most important Greek letter, is the partial derivative of the value of a transaction with respect to the underlying asset price. Traders typically try to maintain delta-neutral (or close to delta-neutral)  positions. For example, the delta of the option mentioned above is about 0.58 so that the delta of the trader's position is $-5,800$ and delta hedging requires 5,800 shares to be purchased as soon as the option is sold. 
 
Because the delta of an option changes during its life, the trader’s position must be rebalanced periodically.  If there are no transaction costs or other frictions, it is in theory optimal to 
 rebalance continuously. If the underlying asset price follows the assumed process, continuous rebalancing will lead to the cost of hedging the option equaling its theoretical price.  In practice, transaction costs and other frictions (which we will collectively refer to as ``trading costs'') mean that the theoretically optimal strategy must be modified. Also, the position in the underlying asset is, in practice, rebalanced periodically rather than continuously. A number of papers have considered the effect of this on option pricing and the efficiency of hedging. These include \cite{leland}, \cite{figlewski}, \cite{boyle}, \cite{grannan}, \cite{toft}, \cite{whalley}, and \cite{martinelli}.

When there are trading costs associated with the underlying asset, delta hedging is no longer optimal. (Indeed, when trading costs are very high it may be optimal to do no hedging at all.) Optimal hedging involves dynamic multi-stage decision making. The decision taken by a trader should take into account the trader’s current holding as well as possible future states of the world and the actions that will be taken in those states. 

To illustrate this, consider again the option mentioned earlier. Assume that trading costs are proportional to the size of a trade. Suppose that after six months the stock price is \$115. The delta of the option is then 0.87.   Delta hedging would require a long position in 8,700 shares, but the costs of trading the stock mean that it may not be optimal to take such a position. A trader who (as a result of earlier hedging) owns 6,000 shares is likely to under-hedge. This under-hedging is likely to be greater than for a trader who already owns 8,000 shares. The trader’s optimal decision in the situation we are considering is influenced by the skewed distribution of future deltas. (If the stock price increases by 10\% during the next month, delta will rise by 0.07 to 0.94; if it decreases by 10\%, it will fall by 0.22 to 0.65.)

This paper examines how reinforcement learning approach can be used to take trading costs into account in hedging decisions. As the example just given would suggest, we find that, when delta hedging would require shares to be purchased, it tends to be optimal for a trader to be under-hedged relative to delta. Similarly, when delta hedging would require shares to be sold, it tends to be optimal for a trader to be over-hedged relative to delta. 
We use an objective function that is the expected cost of hedging plus a constant times the standard deviation of the cost. 

The reinforcement learning approach  can also be used for hedging exposure to volatility. Because the volatility exposure to a derivatives portfolio can be changed only by trading other derivatives, the bid-ask spreads associated with the instruments used for hedging volatility are typically quite high, which increases the relevance of the approach. It is also worth noting that some exotics, such as barrier options, are difficult to hedge using the underlying asset even when there are no trading costs. The approach we illustrate may be useful for these options, even when the trading costs associated with the underlying are negligible.    

Other papers that have used reinforcement learning for hedging include \cite{halperin}, \cite{buehler}, \cite{kolmritter}, and \cite{du}.\footnote{\cite{carbonneau} has tried a similar approach in the context of insurance products.} \cite{halperin} produces an option pricing result when hedging is discrete and there are no trading costs. Kolm and Ritter's work shows how the reinforcement learning approach can be used when the objective function is equal to the mean hedging cost plus a constant times the variance of the hedging cost and certain reasonable assumptions are made to enable the use of a single Q-function. The same approach is used in \cite{du}.  
\cite{buehler} use expected shortfall as an objective function.

We expand the range of objective functions that can be used by extending the standard reinforcement learning algorithms so that both $\EX(C)$ and $\EX(C^2)$ are calculated for different hedging policies where $C$ is the total  future hedging cost during the life of the instrument being hedged.  This idea has been proposed by authors \cite{tamar} in the machine learning literature. We also show how results can be improved by using the deep deterministic policy gradients approach. 

We distinguish two different approaches to calculating hedging costs. Consider again the problem of hedging a short call option position on a stock. In what we refer to as the “accounting P\&L formulation,” the cost of hedging is calculated period-by-period as the change in the value of the hedged position (option plus stock) plus the trading costs associated with changing the position in the stock. In what we refer to as the “cash flow formulation,” the costs are the cash outflows and inflows from trading the stock and there is a potential final negative cash flow payoff on the option.  \cite{kolmritter} use the accounting P\&L formulation while Buehler et al (2019) use the cash flow formulation. The accounting approach requires a pricing model whereas the cash flow approach does not.

The difference between the two approaches concerns the timing of the recognition of gains and losses on the option position. In the accounting P\&L approach, they are recognized period-by-period whereas in the cash flow approach they are recognized at the time of the option payoff. 
We find that the accounting P\&L approach gives much better results than the cash flow approach. This shows that different approaches to decomposing costs can affect the efficiency of learning.\footnote{The reason for our results may be what is referred to as the ``temporal credit assignment problem, '' first discussed by \cite{minsky}} When rewards and costs are measured in terms of P\&L, the results of setting up a particular hedge at the beginning of a period can, to a large extent, be calculated at the end of the period. When cash flows are used, the temporal gap between a hedging decision and its consequences is greater. We have experimented with a hybrid approach where the model used to calculate option prices in the accounting P\&L approach is simpler than the model used to generate the asset prices. We find this works well. It has the advantage that it allows asset price processes with no analytic pricing formulas to be tested. It also corresponds to the way many derivatives desks operate (i.e., they know that the pricing model they use assumes an asset price process that is much simpler than the real-life process).     

One advantage of the hybrid reinforcement learning approach just mentioned is that it can be used in conjunction with any assumption about the process followed by the underlying asset. Ideally, one would like to use market data to train the hedging model. Unfortunately, the amount of data necessary to obtain good results means that this is not feasible. A number of authors have recently suggested interesting approaches for generating synthetic data that is trained to be indistinguishable from market data.\footnote{See, for example, \cite{style}, \cite{generator}, and \cite{wiese}.} 

 Traders would like to find a hedging procedure that works well for a variety of different plausible stochastic processes for the asset price. In order to avoid repeating the analysis for different processes and then averaging the results in some way, a mixture model can be used. Different stochastic processes for the underlying asset  price are specified.  When a path is being simulated for the asset price, one of these processes is randomly selected upfront.  This ensures that the resulting hedging strategy works reasonably well for all the stochastic processes. Reinforcement learning can be a useful tool for this type of analysis even when there are no (or negligible) transaction costs. 

In this paper, we first illustrate the deep hedging approach by assuming that the underlying asset follows geometric Brownian motion. We then assume a stochastic volatility model. In both cases, our reinforcement learning approach outperforms delta hedging when there are trading costs. Furthermore, tests show that, when the asset price process incorporates a stochastic volatility, the results from using a simple Black--Scholes model to revalue options at each step are very similar to those when the stochastic volatility model is used for this purpose.   

The rest of this paper is organized as follows. Section 2 introduces reinforcement learning. Section 3 describes how it can be applied to the hedging decision. Section 4 presents results for the situation where the underlying asset price follows geometric Brownian motion. Section 5 presents results for the situation where the underlying asset follows a stochastic volatility process.  Conclusions are in Section 6.

\section{Reinforcement Learning} \label{rl}

Reinforcement learning considers the situation where a series of decisions have to be made in a stochastically changing environment. At the time of each decision, there are a number of states and a number of possible actions.  The set up is illustrated in Exhibit 1. The decision maker takes an action, $A_0$, at time zero when the state $S_0$ is known.  This results in a reward, $R_1$, at time 1 and a new state, $S_1$, is then encountered.  The decision maker then takes another action, $A_1$ which results in a reward, $R_2$ at time 2 and a new state, $S_2$; and so on.  

\begin{figure}[!t]
    \begin{center}
  {
       \scalebox{0.8}{
      \includegraphics[width=1.00\textwidth]{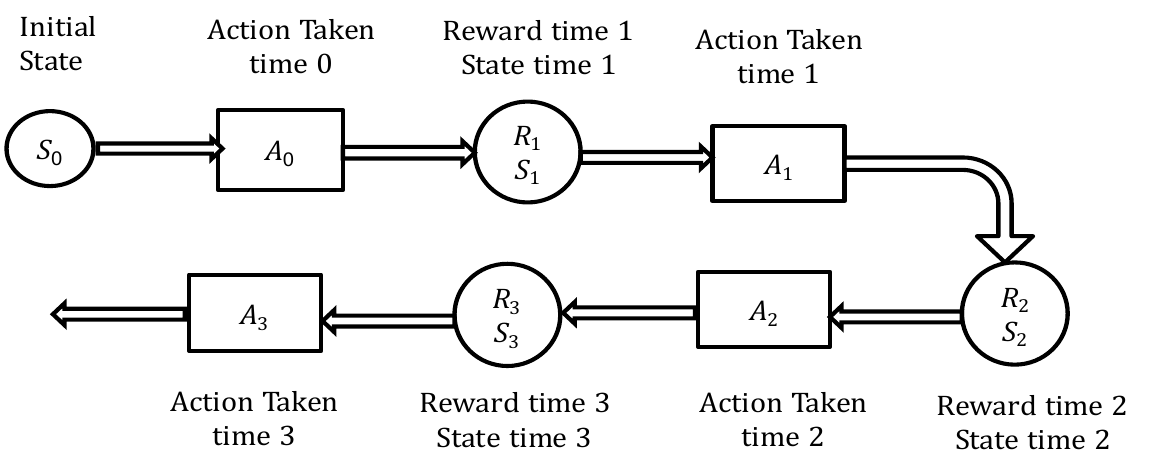}
        }
  }
    \end{center}
    \caption{The  setup in reinforcement learning. $S_i$ is the state at time $i$, $A_i$ is the action taken at time $i$, and $R_{i+1}$ is the resulting reward at time $i+1$.}
    \label{fig:Fig1}
\end{figure}

A comprehensive treatment of reinforcement learning is provided by \cite{sutton}. The aim of reinforcement learning is to maximize expected future rewards. Specifically, it attempts to maximize the expected value of $G_t$ where

\begin{align} \label{returnEq}
G_t = R_{t+1} + \gamma R_{t+2} + \gamma^2 R_{t+3} + \cdots + \gamma^{T-1} R_{T}
\end{align}

\noindent$T$ is a horizon date and $\gamma \in (0,1]$ is a discount factor.  To ensure that  equation (1)  reflects the time value of money, we define $R_t$ as the the cash flow received at time $t$ multiplied by $\gamma$ (i.e., discounted by one period). 

In order to maximize $G_t$ in equation (1), the decision maker needs a set of rules for what action to take in any given state.
This set of rules is represented by a policy function $\pi : \mathcal{S} \rightarrow \mathcal{A}$, where $\mathcal{S}$ and $\mathcal{A}$ are the sets of all possible states and actions, respectively. If the decision maker uses policy $\pi$ and is in state $S_t$ at time $t$, then the action taken is $A_t = \pi(S_t)$. The policy is updated as the reinforcement learning algorithm progresses. As we explain later, learning an optimal policy involves both exploration and exploitation.  For example, at a particular stage in the execution of the reinforcement learning algorithm, the policy might involve, for all states and all times, a 90\% chance of taking the best action identified so far (exploitation) and a 10\% chance of randomly selecting a different action (exploration).

For a specific policy $\pi$, we define the value of taking action $A_t$ in a state $S_t$ as the expected total reward (discounted) starting from state $S_t$ 
taking action $A_t$ and taking the actions given by the policy in the future states that are encountered.
The value of each state-action pair is represented by a function $Q: \mathcal{S} \times \mathcal{A} \rightarrow \mathbb{R}$, referred to as the action-value function or Q-function:

\begin{align} 
\nonumber
Q(S_t, A_t) = \EX(G_t | S_t, A_t) 
\end{align}

The Q-function is estimated from a large number of state and action sequences similar to the one
depicted in Exhibit 1. These are referred to as {\em episodes} and are generated from either historical or simulated data. 
If the decision maker has a good estimate of the value of each action for any state possibly encountered,
then she has an effective way of comparing policies. Reinforcement learning methods provide a way of determining the decision maker's policy as data from episodes are accumulated.

Typically, the decision maker starts with a random policy (100\% exploration) and then gradually increases the amount of exploitation. After each episode of experience, the decision maker observes the returns obtained when choosing certain actions in specific states and updates the Q-function estimates. The decision maker then updates the policy so that it is consistent with the new Q-function estimates. This process is repeated until there is enough confidence that the optimal policy has been trained. 

The estimate of the optimal policy at a particular point during the execution of the reinforcement learning algorithm (which we refer to as the ``current optimal policy'') is the one that takes the highest value action in each state. We denote the current optimal policy by $\pi^*$. When the decision maker is in state $S_t$ the current optimal policy leads to action  $A_t=\pi^*(S_t)=\argmax\limits_{a \in \mathcal{A}} Q(S_t, a)$.
If all state-action pairs have been visited multiple times and convergence has been achieved, then the current optimal policy can
be assumed to be the true optimal policy. 

Directly using $\argmax()$ to determine a policy is suitable when the action space is discrete and not too large. However, when the action space is continuous, this method becomes impractical as it involves a global maximization at every step. An alternative approach, known as {\em policy gradients}, learns the optimal policy directly. The basic idea is to represent the policy by a parametric function $\pi(S_t; \theta)$, where $\theta$ is the parameter vector. Learning is then done through exploring the environment and adjusting the policy parameters in the direction of greater action values. 

In what follows, we discuss techniques to handle the two core stages of the reinforcement learning process: the estimation of the Q-function and the policy update. There are many approaches. We will focus on the ones that are relevant to this paper: Monte Carlo (MC) methods, temporal difference (TD) learning, and policy gradients (PG).  They offer different trade-offs with respect to convergence speed, bias, and variance of the estimated values.

\subsection{Monte Carlo}

MC methods estimate the value of state-action pairs by computing the mean return from many episodes. The updating rule is

\begin{align} 
\label{MCupdate}
Q(S_t,A_t) \leftarrow Q(S_t,A_t) + \alpha\Big{(}G_t - Q(S_t,A_t)\Big{)} 
\end{align}

\noindent where $\alpha \in (0,1]$ is a constant parameter. 
Before performing an update, MC methods need to wait until an episode is completed, since the update rule requires $G_t$, which sums over discounted future rewards from time $t+1$ to time $T$ when the episode ends. This simple approach however provides an unbiased estimator of the true action values if data on enough 
episodes are available.

\subsection{Temporal Difference Learning}

TD methods also estimate the Q-function from many episodes. They do this by looking ahead to the next time step.
Specifically, at time $t$ the policy at time $t$ leads to a reward $R_{t+1}$ at time $t+1$ and the Q-function is updated as follows:

\begin{align} 
\label{tdupdate}
Q(S_t,A_t) \leftarrow Q(S_t,A_t) + \alpha \Big{(}R_{t+1} + \gamma Q(S_{t+1}, A_{t+1}) - Q(S_t,A_t)\Big{)} 
\end{align}

\noindent where $\alpha \in (0,1]$ is  a constant parameter and $\gamma$ is the discount factor introduced earlier.\footnote{Other variants of TD apply updates by looking more than one time step ahead as described by \cite{sutton}.} 

TD updates exhibit less variance than MC updates since the only source of uncertainty in the update at time $t$ comes from a single time step ahead in the episode, rather than from the entire sequence $t+1, \dots, T$.  
It should be noted, however, that TD estimates are more sensitive to the initialization of the Q-function, since any error in the current estimate influences the next TD update. If the initial Q-function values are far from ground truth, then TD approximations are generally biased. \cite{kearns} have shown that error bounds in TD exponentially decay as more information is accumulated from episodes.   

What is common to the above MC and TD approximations is that they are {\em on-policy}: the values updated correspond to
the current policy that the decision maker is following. For example, if the decision maker is following a random policy to
explore the environment, then the state-action values represent the returns that the decision maker is expected to
receive if she follows exactly that random policy. The aim of course is to update the policy in a way that ensures it converges to the optimal policy.

\subsection{Off-policy TD: Q-Learning} 

\cite{watkins} introduced Q-learning, a TD method that is {\em off-policy}, meaning that the decision maker can always update the value estimates of the optimal policy while following a possibly sub-optimal policy that
permits exploration of the environment.

In Q-learning, the Q-function updates are decoupled from the policy currently followed and take the following form:

\begin{align} 
Q(S_t,A_t) \leftarrow Q(S_t,A_t) + \alpha \Big{(}R_{t+1} + \gamma \max\limits_{a \in \mathcal{A}} Q(S_{t+1}, a) - Q(S_t,A_t)\Big{)} 
\end{align}

\begin{figure}
    \fontsize{7}{9}\selectfont
    \def\svgwidth{.48\textwidth}
    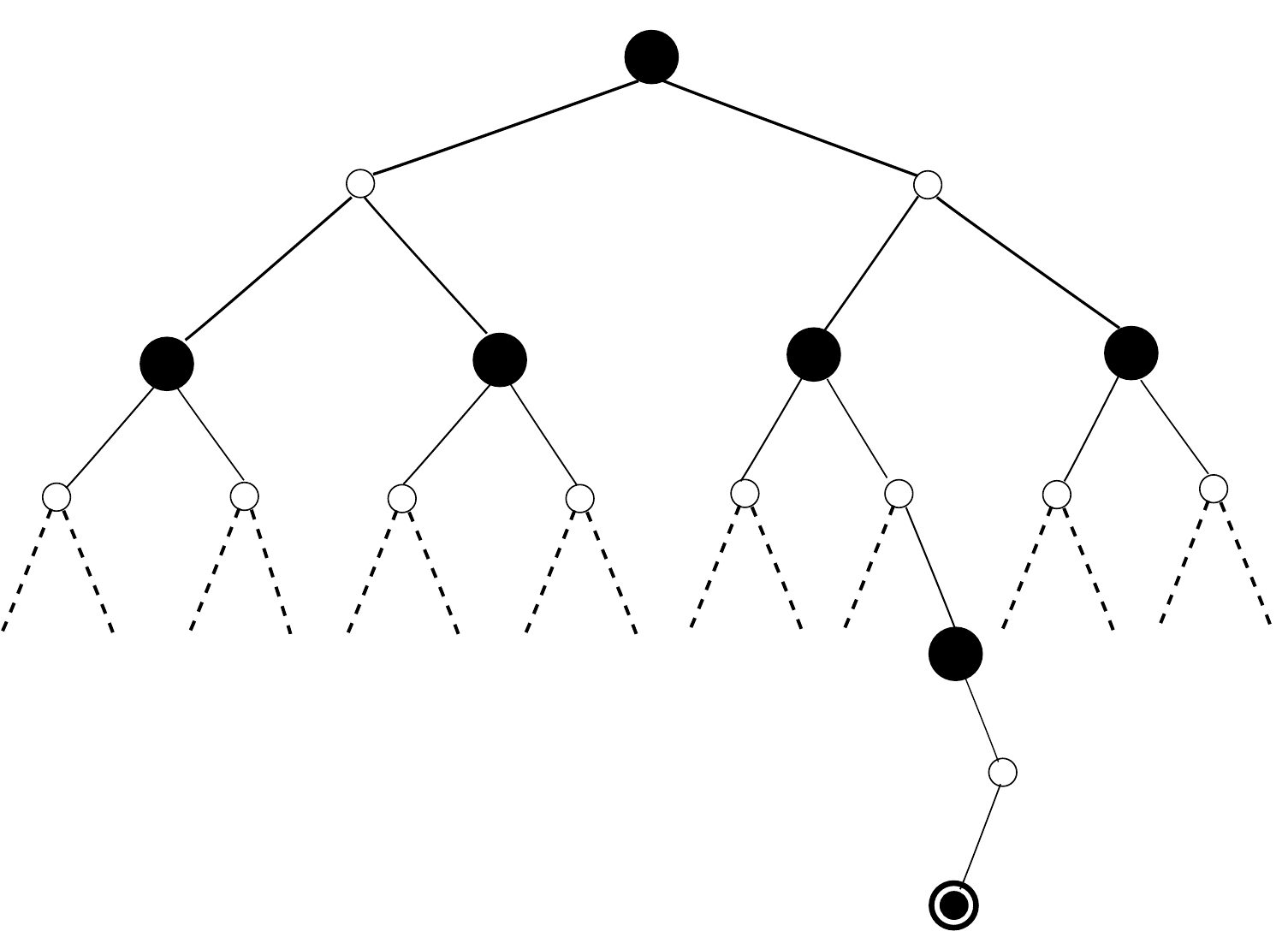
    \hfill
    \fontsize{7}{9}\selectfont
    \def\svgwidth{.48\textwidth}
    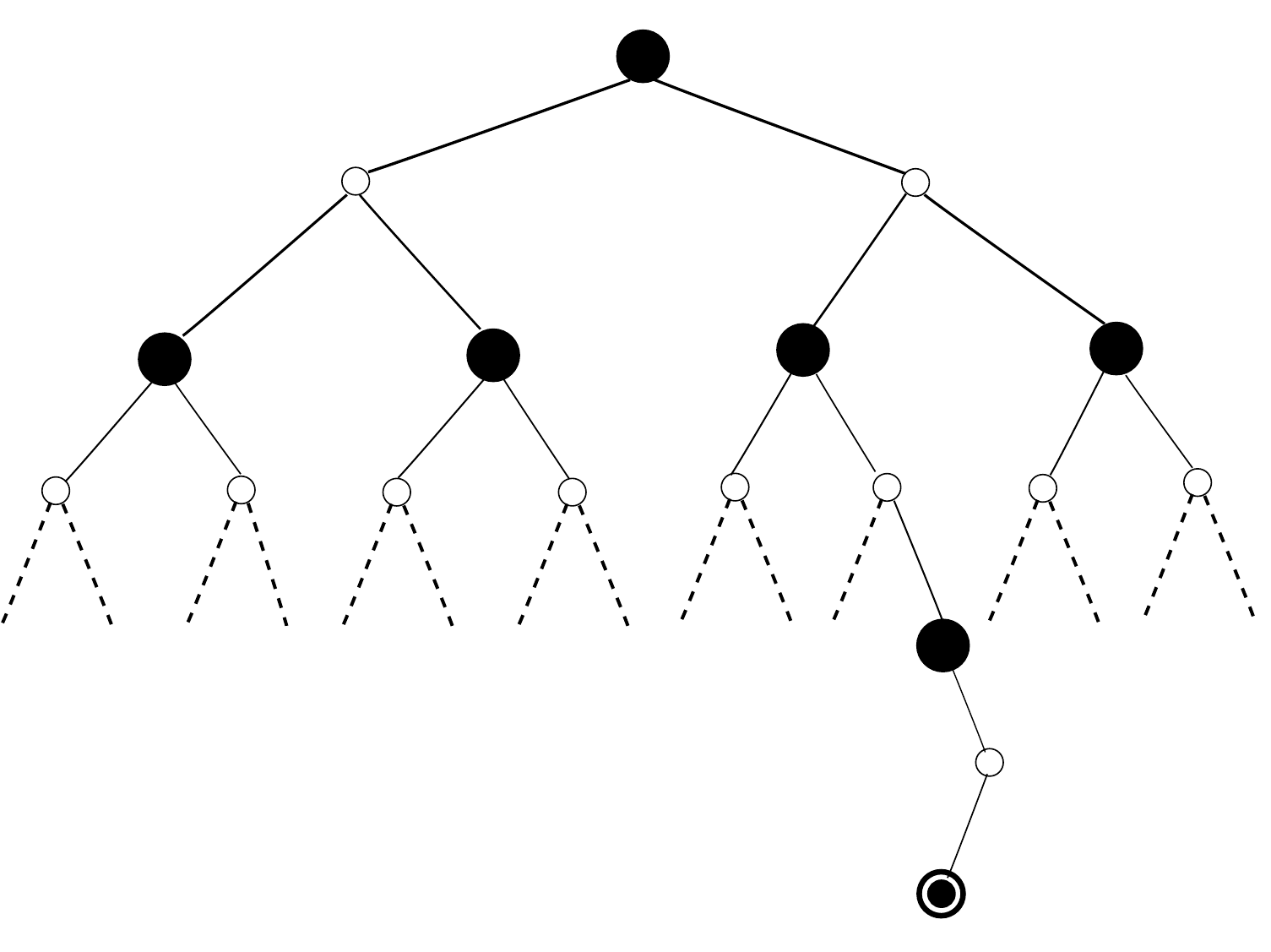
    \caption{Tree representation of action-value search space. TD updates only require action values that are one time step ahead from $S_t$, while MC methods require values for the entire sequence $t+1,\dots, T$.}
    \label{fig:Fig2}
\end{figure}

The decision maker enters state $S_t$, chooses action $A_t$ based on the currently followed policy $\pi$, observes the reward $R_{t+1}$ and updates the value $Q(S_t,A_t)$ according to what the expected returns are if the current optimal policy $\pi^{\star}$ is followed at state $S_{t+1}$. 
Convergence guarantees for Q-learning exist as long as state-action pairs continue to be visited by the decision maker's policy
while the action values of the optimal policy are updated.

Exhibit 2 illustrates how the dependencies between state transitions differ when MC or TD updates are performed for a single state-action pair.

\subsection{Policy Update} \label{sec:egreedy}

Every time the Q-function is updated either by MC, TD or Q-learning methods, the decision maker can adjust the policy 
based on the new value estimates. The current optimal policy $\pi^{\star}$, as mentioned earlier, involves choosing the action in each state that leads to the maximum Q-value. This is referred to as the {\em greedy} policy. 

The decision maker cannot merely follow the greedy policy after each update because it might not be a good estimate of the optimal policy when the Q-function has not converged to its true value. Always following the greedy policy would lead to no exploration of other state-action pairs. Instead, the decision maker follows what is termed  an $\epsilon$-greedy policy, denoted $\pi_{\epsilon}$. this acts like a  greedy policy with some probability $1-\epsilon$
and resembles a random policy with probability $\epsilon$:

\begin{equation}
\nonumber
\pi_{\epsilon}(S_t) = 
\begin{cases}
\argmax\limits_{a \in \mathcal{A}} Q(S_t, a), & \text{with probability $1-\epsilon$} \\
\text{random $a \in \mathcal{A}$}, & \text{with probability $\epsilon$}
\end{cases}
\end{equation}

The probability $\epsilon$ of selecting a random action starts high (often $\epsilon = 1$) and allows the decision maker to explore 
multiple states and actions, which in turn leads to estimates for a wide range of state-action values. As learning progresses, $\epsilon$ slowly decays towards
zero. The decision maker starts following the greedy policy some of the time, while still exploring random alternatives. Later in the process, the greedy 
policy dominates and is almost always followed. The final greedy policy is the algorithm's estimate of the optimal policy.

\subsection{Deep Q-learning}

When there are many states or actions (or both), a very large number of episodes can be necessary to provide sufficient data on the state-action combinations. A way of handling this problem is to use an artificial neural network (ANN) to estimate a complete Q-function from the results that have been obtained, as proposed by~\cite{mnih2015}. This approach allows the state space to be continuous.

Using Q-learning in conjunction with an ANN is referred to as deep Q-learning or deep reinforcement learning.
In this setting, the ANN's estimation of the Q-function for state-action pair $(S_t,A_t)$ is denoted as $Q(S_t,A_t; \theta)$, where $\theta$ are the ANN's parameters.
The goal is to develop a process where the parameters are iteratively updated so as to minimize the error between the estimated $Q(S_t,A_t; \theta)$ and the true Q-function $Q(S_t,A_t)$. However, since the true Q-function is not known, the parameters are instead adjusted to minimize the error between the 
ANN's current estimation of $Q(S_t,A_t; \theta)$ and what the estimation should be if it is updated using Q-learning, as in Section 2.3.
For a single state-action pair that is observed in a collection of episodes, it is common to minimize the squared error

\begin{align} 
\nonumber
\Big{(}R_{t+1} + \gamma \max\limits_{a \in \mathcal{A}} Q(S_{t+1}, a) - Q(S_t,A_t;\theta)\Big{)}^2 
\end{align}

The ANN's parameters can be updated via gradient descent. The process repeats for all state-action pairs that have been collected.
To stabilize the learning process, the network's parameters are often updated
by measuring the error over sets of state-action pairs, referred to as batches. 
Typical sizes for a batch range from $10$ to the thousands, depending on the application.

Using ANNs in reinforcement learning introduces additional challenges. Most ANN learning algorithms assume that the samples used for training are independently and identically distributed. When the samples are generated from sequentially exploring the environment, this assumption no longer holds. To mitigate this issue, deep Q-learning algorithms use a replay buffer, where sequential samples are first stored in the buffer, and then randomly drawn in batches to be used in training. This technique of removing the correlations between sequential samples is referred to as experience replay. Some experiences are more valuable for learning than others. \cite{replay} proposes a method for prioritizing experience, where experiences from which there is more to learn are replayed more often. Prioritized experience replay improves data efficiency and often leads to faster learning.

Another challenge is that in standard Q-learning the update target of the current Q-function is constructed using the current Q-function itself. This leads to correlations between the current Q-function and its target, and in turn destabilizes learning. To improve stability, deep Q-learning keeps a separate copy of the Q-function for constructing the update target, and only updates this copy periodically.

\subsection{Deterministic Policy Gradient (DPG) and Deep DPG}

The policy update step of the Q-learning method entails a global optimization (the $argmax()$ operation) at each time step. When the action space is continuous, this maximization becomes intractable. Deterministic policy gradient (DPG), an algorithm proposed by \cite{dpg}, avoids the costly update by learning the policy function directly.

Specifically, DPG parameterizes the action-value function and policy function as $Q(S_t, A_t; w)$ and $\pi(S_t; \theta)$, where $w$ and $\theta$ are ANN parameter vectors. Since the policy is deterministic the decision maker follows an $\epsilon$-greedy policy similar to the one in section~\ref{sec:egreedy} to ensure exploration during the learning. With probability $\epsilon$, a random action is taken, and with probability $1-\epsilon$ the policy function is followed.

The learning of $Q$ is similar to the one in Deep Q-learning, an iterative process of updating the parameter vector $w$ via gradient decent to minimize the squared error of the current Q and it's target update value:

\begin{align} 
\nonumber
\Big{(}R_{t+1} + \gamma Q(S_{t+1}, \pi(S_{t+1})) - Q(S_t, A_t; w)\Big{)}^2 
\end{align}

To update the policy, instead of finding the action that maximizes the $Q$-function, a gradient ascent algorithm is employed to adjust the parameter $\theta$ in the direction of the gradient of $Q(S_t, \pi(S_t; \theta)$, i.e. in the direction of the fastest increase of $Q$-function: 

\begin{align} 
\nonumber
\theta \leftarrow \theta + \alpha \nabla_{\theta}Q(S_t, \pi(S_t; \theta)) 
\end{align}

In other words, the policy is updated at each step to return higher action values. \cite{dpg} present a detailed analysis proving that the above parameter update leads to locally optimal policies.

The policy function and $Q$-function in the DPG method are referred to as actor and critic, respectively. The actor decides which action to take, and the critic evaluates the action and updates the actor so that better actions are taken in subsequent steps. The actor and critic repeatedly interact with each other until convergence is achieved.

Deep DPG suggested by \cite{lillicrap} combines the ideas of DPG and Deep Q-learning. Using ANNs as function approximators for both the policy and action-value functions, deep DPG follows the basic algorithm outlined above and addresses the challenges of training ANNs as in Deep Q-learning. 

\section{Application to Hedging} \label{hedging}

The rest of this paper focuses on the application of reinforcement learning to hedging decisions. A stochastic process for the underlying asset is specified and episodes are generated by simulating from this stochastic process.

We use as an example the situation where a trader is hedging a short position in a call option. We assume that the trader rebalances her position at time intervals of $\mathrm{\Delta}
t$ and is subject to trading costs. The life of the option is $n \mathrm{\Delta} t$. The cost of a trade in the underlying asset in our formulation is proportional to the value of what is being bought or sold, but the analysis can easily be adjusted to accommodate other assumptions. The state at time $i\mathrm{\Delta} t$  is defined by three parameters:
\begin{enumerate}
\item The holding of the asset during the previous time period; i.e., from time $(i-1) \mathrm{\Delta} t$ to time $i\mathrm{\Delta} t$
\item The asset price at time $i\mathrm{\Delta} t$
\item The time to maturity
\end{enumerate}
The action at time $i\mathrm{\Delta} t$ is the amount of the asset to be held for the next period; i.e., from time $i\mathrm{\Delta} t$ to time $(i+1)\mathrm{\Delta} t$. 

 There are two alternative formulations of the  hedger's problem: the accounting P\&L formulation and the cash flow formulation. For ease of exposition we assume that $\gamma=1$ (no discounting).

\subsection{Accounting P\&L formulation}

In the accounting P\&L formulation,  rewards (negative costs) are given by  
\begin{align}
\nonumber
R_{i+1}= V_{i+1}-V_i+H_i(S_{i+1}-S_i)-\kappa\vert S_{i+1}(H_{i+1}-H_i)\vert
\end{align}
for $0\le i<n$ where $S_i$ is the asset price\footnote{For the rest of the paper $S$ may refer to the asset price as opposed to the state, 
depending on the context it is used in.} at the beginning of period $i$, $H_i$ is the holding between time $i\mathrm{\Delta} t$ and $(i+1)\mathrm{\Delta} t$,  $\kappa$ is the trading cost as a proportion of the value of what is bought or sold, and $V_i$ is the value of the derivative position at the beginning of period $i$. ($V_i$ is negative in the case of a short call option position.) In addition, there is an initial reward associated with setting up the hedge equal to $-\kappa\vert S_0H_0\vert$ and a final reward associated with liquidating the hedge at the end equal to $-\kappa\vert S_nH_{n}\vert)$.

\subsection{Cash Flow Formulation}

In the cash flow formulation the rewards are given by
\begin{align} 
\nonumber
R_{i+1}=S_{i+1}(H_{i}-H_{i+1})-\kappa\vert S_{i+1}(H_{i+1}-H_{i})\vert
\end{align}
for $0\le i<n$. There is an initial cash flow associated with setting up the hedge equal to $-S_0H_0-\kappa\vert S_0H_0\vert$. At the end of the life of the option there is a final negative cash flow consisting of (a) the liquidation of the final position (if any) in the underlying asset which equals $S_nH_{n}-\kappa\vert S_nH_{n}\vert$ and (b) the payoff (if any) from the option. Note that the cash flow formulation requires the decision maker to specify a stochastic process for the underlying asset but (unlike the accounting P\&L formulation) it does not require her to specify a pricing model. The algorithm in effect needs to learn the correct pricing model. The formulation allows the decision maker to use stochastic processes for which there are no closed form pricing models.\footnote{Note that, whereas a perfect hedge in the accounting P\&L formulation will give rise to zero reward in each period, it will give rise to positive and negative rewards in each period in the  cash flow formulation. The total reward, including the initial cash flow received for the option, will be zero.} 

\subsection{Hybrid Approach}

As mentioned earlier, we find that the accounting  P\&L approach gives much better results than the cash flow approach (possibly because of a temporal credit assignment problem). We find that a hybrid approach, where the model used to value the option is simpler than the model used to generate asset prices, works well. In this context, it is worth noting that on any trial the total cost of hedging an option (assuming no discounting) is independent of the option pricing model used. The hybrid approach does not therefore bias results. Its objective is simply to use a plausible pricing model that reduces the impact of temporal differences between actions and outcomes. 

\subsection{Our Set Up}

In the problem we are considering, it is natural to work with costs (negative rewards). This is what we will do from now on.   We use an objective function which is the expected hedging cost plus a constant multiplied by the standard deviation of the hedging cost. Define
\begin{align}
Y(t)=\EX(C_t) + c\sqrt{\EX(C_t^2)-\EX(C_t)^2} 
\end{align}

\noindent where $c$ is a constant and $C_t$ is the total hedging cost from time $t$ onward. Our objective is to minimize $Y(0)$. We assume that the decision maker pre-commits to using $Y(t)$ as the objective function at time $t$ for all $t$. The objective function does have some attractive properties. It is a coherent risk measure.\footnote{See~\cite{risk} for a definition and discussion of coherence.}  It also satisfies the Bellman equation.\footnote{This is because, when costs of $A$ have been incurred between time zero and time $t$, $\min Y(0)= A + \min Y(t)$, where minimization is taken over all actions.}    

As explained below, we accommodate the objective function in equation (5) by defining two Q-functions . The first Q-function, $Q_1$, estimates the expected cost for state-action combinations. The second Q-function, $Q_2$ estimates the expected value of the squared cost for state-action combinations. Other objective functions, $Y(t)$, can be accommodated similarly by defining appropriate Q-functions. For example, if the objective function involves skewness so that downside risk is penalized more than upside potential, an additional Q-function could be defined to monitor the third moment of the cost.

The Q-learning algorithm in Section 2.3 can be adapted for this problem if we discretize the action space, for example, by rounding the hedge in some way. The algorithm proceeds as described in the previous section except that the greedy action, $a$, is the one that minimizes

\begin{equation}
F(S_t, a)=Q_1(S_t, a)+c\sqrt{Q_2(S_t, a)-Q_1(S_t, a)^2)}
\end{equation}

From equation (4), the updating formula for $Q_1$ is

\begin{equation}
\nonumber
Q_1(S_t,A_t) \leftarrow Q_1(S_t,A_t) + \alpha \Big{(}R_{t+1} + \gamma Q_1(S_{t+1}, a) - Q_1(S_t,A_t)\Big{)} 
\end{equation}

\noindent where the action, $a$, is the greedy action that minimizes $F(S_{t+1}, a)$. The expected value of $Q_2(S_t,A_t)$ if the greedy action is followed is the expected value of $[R_{t+1}+\gamma Q_1(S_{t+1}, a)]^2$. This means that the updating rule for $Q_2(S_t,A_t)$ is: 
\begin{equation}
\nonumber
 Q_2(S_t,A_t) \leftarrow Q_2(S_t,A_t) + \alpha \{R_{t+1}^2 + \gamma^2  Q_2(S_{t+1}, a) +2\gamma R_{t+1}Q_1(S_{t+1}, a) - Q_2(S_t, A_t)\}
\end{equation}

\noindent where action $a$ is the same greedy action as that used to update $Q_1$.

Similarly, if we take a continuous action space approach, the DPG (or deep DPG) algorithm can be adapted to estimate the two Q-functions and update the policy function. The loss functions needed for learning the parameterized $Q_1(S_t, A_t; w_1)$ and $Q_2(S_t, A_t; w_2)$ functions are

\begin{gather}
\nonumber
\Big{(}R_{t+1} + \gamma Q_1(S_{t+1}, \pi(S_{t+1})) - Q_1(S_t, A_t; w_1)\Big{)}^2,\\
\nonumber
\Big{(}R_{t+1}^2 + \gamma^2  Q_2(S_{t+1}, \pi(S_{t+1})) +2\gamma R_{t+1}Q_1(S_{t+1}, \pi(S_{t+1})) - Q_2(S_t, A_t; w_2)\Big{)}^2.
\end{gather}

The update of policy function $\pi(S_t; \theta)$ follows

\begin{align} 
\nonumber
\theta \leftarrow \theta - \alpha \nabla_{\theta}F(S_t, \pi(S_t; \theta)).
\end{align}

\section{Geometric Brownian Motion Test} \label{results}

As a first test, we assume that the stock price, $S$, follows geometric Brownian motion:
\begin{equation} \label{qval}
\nonumber
dS=\mu S dt+\sigma S dz\\
\end{equation}

\noindent where $\mu$ and $\sigma$ are the stock's mean return and volatility (assumed constant), and $dz$ is a Wiener process. The work of \cite{bs} and \cite{merton} show that the price of a European call option with maturity $T$ and strike price $K$ is 

\begin{equation} 
S_0e^{-qT}N(d_1)-Ke^{-rT}N(d_2)\\
\end{equation}

\noindent where $r$ is the risk-free rate, $q$ is the dividend yield (both assumed constant), $S_0$ is the initial value of $S$ and 

\begin{align*}
\nonumber
d_1&={{\ln(S_0/K)+(r-q+\sigma^2/2)T}\over{\sigma\sqrt{T}}}\\
\nonumber
d_2&=d_1-\sigma\sqrt{T}
\end{align*}

We choose to solve the reinforcement learning problem using the deep DPG method as the method allows the hedging position to be continuous. (Unlike the Q-learning method, it does not require a discrete set of hedging positions in the underlying asset to be specified.) To improve data efficiency and learning speed, we also implement the prioritized experience replay method. As indicated earlier, the  accounting P\&L approach gives better results than the cash flow approach.\footnote{We also tested Q-learning method, under which we discretize the action space by rounding the hedge position to the nearest 10\% of the assets underlying the option. Again the accounting P\&L approach gives better results than the cash flow approach.}  This may be related to the credit assignment problem, details of which can be found in the work by \cite{minsky}. In broad terms, it is challenging to match the consequences of an action to the rewards the decision maker receives in the cash flow approach. The decision maker must examine the rewards over long time periods to get necessary information and, as a result, learning is more difficult. The reward stream obtained using the cash flow approach often consists of rewards that exhibit relatively high volatility and for which an immediate relation to what would constitute a good action is hard to infer. In the accounting P\&L approach pricing information is implicitly provided to the model. Thus, rewards that are associated with actions leading to a perfect hedge are closer to zero, and this is informative on a per-period basis. In the cash flow set up, on the other hand, the correct pricing model needs to be ``discovered" by the learning algorithm at the same time as the optimal policy is searched for. This interdependence renders the reinforcement learning algorithms more sensitive to hyper-parameters and initialization methods. In what follows, all results were produced using the accounting P\&L approach.

Exhibits 3 and 4 compare the results from using reinforcement learning with delta hedging for short positions in at-the money ($S_0=K$) call options on a stock lasting one month and three months when  $\mu =5\%$, $r=0$, $q=0$.\footnote{Note that although the price of the option does not depend on $\mu$, the results from using a particular hedging policy are liable to do so.} We set $c=1.5$ in equation (5) so that the hedger's objective is to minimize the mean cost of hedging plus 1.5 times the standard deviation of the cost of hedging. The trading cost parameter, $\kappa$, is 1\%.

\begin{table}[]
\centering

\begin{tabular}{|c|c|c|c|c|c|}
\hline
 &  \multicolumn{2}{|c|}{Delta Hedging} & \multicolumn{2}{c|}{RL Optimal Hedging} & $Y(0)$ \\ 
\cline{2-5}
 Rebal Freq & Mean Cost & S.D. Cost & Mean Cost  & S.D. Cost & improvement \\ \hline
weekly & 69\% & 50\% & 60\% & 54\% & 1.7\% \\
3 days & 78\% & 42\% & 62\% & 48\% & 4.7\% \\
2 days & 88\%  & 39\% & 73\% & 41\% & 8.5\% \\
daily & 108\% & 38\% & 74\% & 42\% & 16.6\% \\ \hline
\end{tabular}%

\bigskip
\caption{Cost of hedging a short position in a one-month at-the-money call option as a percent of the option price when the trading cost is 1\%. The last column shows the performance of RL hedging versus Delta hedging expressed as the percentage improvement with respect to the objective function $Y(0)$: $(Y(0)_{Delta} - Y(0)_{RL}) / Y(0)_{Delta}$. Asset price follows geometric Brownian motion with 20\% volatility. The (real-world) expected return on the stock is 5\%. The dividend yield and risk-free rate are zero.}
\label{Table:Tab1}
\end{table}

\begin{table}{}
\centering

\begin{tabular}{|c|c|c|c|c|c|}
\hline
 &  \multicolumn{2}{|c|}{Delta Hedging} & \multicolumn{2}{c|}{RL Optimal Hedging} & $Y(0)$ \\ 
\cline{2-5}
 Rebal Freq & Mean Cost & S.D. Cost & Mean Cost  & S.D. Cost & improvement \\ \hline
weekly & 55\% & 31\% & 44\% & 38\% & 0.2\% \\
3 days & 63\% & 28\% & 46\% & 32\% & 10.9\% \\
2 days & 72\%  & 27\% & 50\% & 29\% & 16.6\% \\
daily & 91\% & 29\% & 53\% & 28\% & 29.0\% \\ \hline
\end{tabular}%

\bigskip
\caption{Cost of hedging a short position in a three-month at-the-money call option as a percent of the option price when the trading cost is 1\%. The last column shows the performance of RL hedging versus Delta hedging expressed as the percentage improvement with respect to the objective function $Y(0)$: $(Y(0)_{Delta} - Y(0)_{RL}) / Y(0)_{Delta}$. Asset price follows geometric Brownian motion with 20\% volatility. The (real-world) expected return on the stock is 5\%. The dividend yield and risk-free rate are zero.}
\label{Table:Tab2}
\end{table}

Exhibits 3 and 4 show that using RL optimal hedging rather than delta hedging has small negative effect on the standard deviation of the cost of hedging in the situations we consider, but  markedly improves the mean cost of hedging. In the case of the one-month option, the mean cost of daily hedging is reduced by about 31\% while in the case of the three-month option it is reduced by about 42\%. Overall, as shown in the last columns of Exhibits 3 and 4, in terms of our optimization objective $Y(0)$, RL optimal hedging outperforms delta hedging in all cases we consider. The percentage improvement of RL hedging over delta hedging increases as rebalancing becomes more frequent. As the life of the option increases the cost of hedging as a percent of the price of the option declines while the gain from replacing delta hedging by an optimal strategy increases.

\begin{figure}[!t]
    \begin{center}
  {
       \scalebox{0.8}{
      \includegraphics[width=1.00\textwidth]{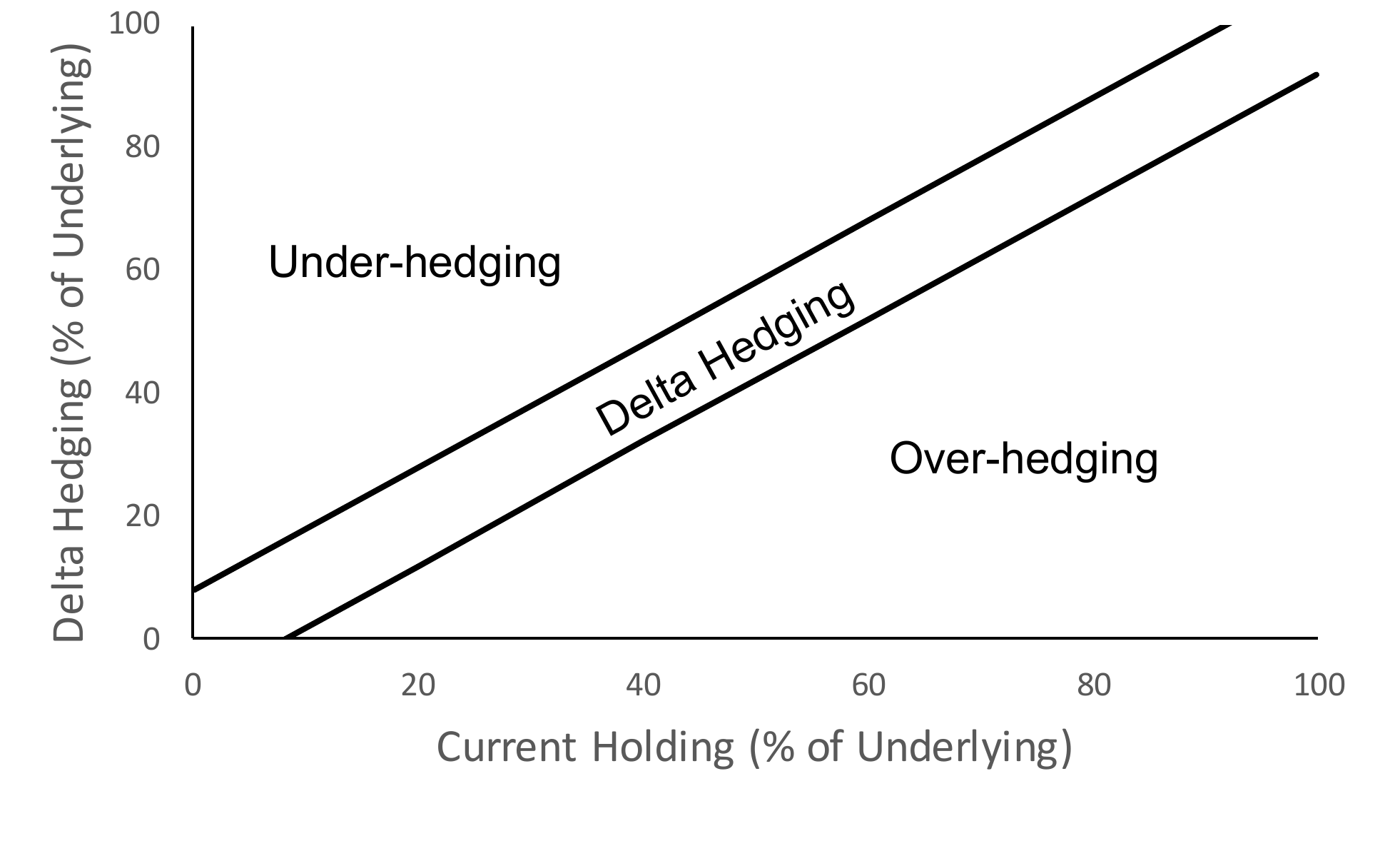}
        }
  }
    \end{center}
    \caption{Over-hedging and under-hedging relative to delta hedging when the optimal policy is adopted in the presence of transaction costs.}
    \label{fig:Fig3:}
\end{figure}

Whereas the performance of delta hedging gets progressively worse as the frequency of hedging increases, the optimal hedging strategy should get progressively better. For example, hedging once a day should give a result at least as good as hedging once every two days because the second strategy is a particular case of the first strategy. Due to the stochastic nature of the learning algorithm it is notable that RL may not always lead to improvements in the objective function as the rebalancing frequency is increased. For example, this can be observed when contrasting the RL costs between the two-day and one-day strategies. Despite the limitation, the RL method consistently outperforms delta hedging with an improvement gap that becomes wider as the rebalancing frequency increases.

An inspection of the decisions taken by the reinforcement learning model shows that they correspond to the policy mentioned earlier. This policy is illustrated in Exhibit 3. When the current holding is close to the holding required for delta hedging, it is optimal for the trader to be close-to-delta hedged. When the holding is appreciably less than that required for delta hedging it is optimal for the trader to be under-hedged (relative to delta). When the holding is appreciably more than that required for delta hedging it is optimal for the trader to be over-hedged (relative to delta).

\section{Stochastic Volatility Test}

As a second test of the reinforcement learning approach, we assume an extension of geometric Brownian motion where the volatility is stochastic:

\begin{align*} 
\nonumber
dS&=\mu S dt+\sigma S dz_1\\
\nonumber
d\sigma&=v\sigma dz_2
\end{align*}

In this model $dz_1$ and $dz_2$ are two Wiener processes with constant correlation $\rho$ and $v$ (a constant) is the volatility of the volatility. The initial value of the volatility $\sigma$ will be denoted by $\sigma_0$. The model is equivalent to a particular case of the SABR model developed by Hagen et al (2002) where the parameter, usually denoted by $\beta$ in that model, is set equal to one.\footnote{The general SABR model is $dF=\sigma F^{\beta}dz_1$ with $d\sigma=v\sigma dz_2$ where $F$ is the forward price of the asset for some maturity. We assume $r$ and $q$ are constant and $\beta=1$ to create a model for $S$ that is a natural extension of geometric Brownian motion.} 
Defining

\begin{align*} 
\nonumber
F_0&=S_0e^{(r-q)T}\\
\nonumber
B&=1+\left({{\rho v\sigma_0}\over {4}}+{{(2-3\rho^2)v^2}\over {24}}\right)T\\
\nonumber
\phi&={{v}\over{\sigma_0}}\ln\left({{F_0}\over{K}}\right)\\
\nonumber
\chi&=\ln\left({{\sqrt{1-2\rho\phi+\phi^2}+\phi-\rho}\over{1-\rho}}\right)
\end{align*}

\noindent Hagen et al show that the implied volatility is approximately $\sigma_0 B$ when $F_0=K$ and $\sigma_0 B \phi/\chi$ otherwise. When substituted into equation (7) the implied volatility gives the price of the option. 

We assume that only the underlying asset is available for hedging the option. 
A popular hedging procedure, which we refer to as ``practitioner delta hedging" involves using a delta calculated by assuming the Black--Scholes model in equation (7) with $\sigma$ set equal to the current implied volatility.\footnote{For a European call option this delta is, with the notation of equation (7), $e^{-qT}N(d_1)$.} \cite{bartlett} provides a better estimate of delta for the SABR model by considering both the impact of a change in $P$ and the corresponding expected change in $\sigma$. This has become known as ``Bartlett's delta."

Exhibits 6 and 7 show the standard deviation of the cost of hedging a one- and three-month option as a percent of the option price when practitioner delta hedging, Bartlett delta hedging, and optimal hedging, as calculated using reinforcement learning, are used. We assume that the initial volatility is 20\% and that $\rho=-0.4$, $v=0.6$, and $\sigma_0=20\%$. The values of $r$, $q$, $\mu$, and $c$ are the same as in the geometric Brownian motion case. The results are remarkably similar to those for geometric Brownian motion. In the absence of trading costs it is well known that (a) delta hedging works noticeably less well in a stochastic volatility environment than in a constant volatility environment and (b) Bartlett delta hedge works noticeably better that practitioner delta hedging in a stochastic volatility environment. These results do not seem to carry over to a situation where there are large trading costs.

\begin{table}[!h]
\centering
\begin{tabular}{|c|c|c|c|c|c|c|c|c|}
\hline
& \multicolumn{2}{|c|}{Bartlett Delta} & \multicolumn{2}{c|}{Practitioner Delta} & \multicolumn{2}{c|}{RL Optimal} 
& $Y(0)$ improv. & $Y(0)$ improv. \\
\cline{2-7}
Rebal Freq & Mean & S.D. & Mean & S.D. & Mean & S.D. & vs. Bartlett & vs. Delta\\ 
\hline
weekly & 69\% & 51\% & 69\% & 50\% & 56\% & 57\% & 2.6\% & 1.8\% \\
3 days & 78\% & 44\% & 78\% & 43\% & 61\% & 51\% & 4.5\% & 3.5\% \\
2 days & 88\%  & 41\% & 88\% & 40\% & 62\% & 52\% & 6.9\% & 6.0\% \\
daily & 108\% & 39\% & 108\% & 38\% & 71\% & 45\% & 16.7\% & 15.9\% \\ \hline
\end{tabular}%

\bigskip
\caption{Cost of hedging a short position in a one-month option as a percent of the option price when the asset price follows the SABR model with $\beta=1.0$, and $\rho=-0.4$. The initial volatility is 20\% and the volatility of the volatility is 60\%. The real world expected return on the stock is 5\%. The dividend yield and risk-free rate are zero. The last two columns show the performance of RL hedging versus Bartlett delta and practitioner delta hedging expressed as the percentage improvement with respect to the objective function $Y(0)$.}
\label{Table:Tab3}
\end{table}

\begin{table}[!h]
\centering
\begin{tabular}{|c|c|c|c|c|c|c|c|c|}
\hline
& \multicolumn{2}{|c|}{Bartlett Delta} & \multicolumn{2}{c|}{Practitioner Delta} & \multicolumn{2}{c|}{RL Optimal} 
& $Y(0)$ improv. & $Y(0)$ improv. \\
\cline{2-7}
Rebal Freq & Mean & S.D. & Mean & S.D. & Mean & S.D. & vs. Bartlett & vs. Delta\\ 
\hline
weekly & 55\% & 36\% & 55\% & 35\% & 42\% & 43\%  & 2.5\% & 0.5\% \\
3 days & 64\% & 33\% & 64\% & 32\% & 48\% & 39\% & 7.3\% & 5.3\% \\
2 days & 72\% & 33\% & 72\% & 31\% & 54\% & 34\% & 13.7\% & 11.9\% \\
daily  & 91\% & 35\% & 91\% & 33\% & 46\% & 38\% & 27.9\% & 26.4\% \\ \hline
\end{tabular}%

\bigskip
\caption{Cost of hedging a short position in a three-month option as a percent of the option price when the asset price follows the SABR model with $\beta=1.0$ and $\rho=-0.4$. The initial volatility is 20\% and the volatility of the volatility is 60\%. The real world expected return on the stock is 5\%. The dividend yield and risk-free rate are zero. The last two columns show the performance of RL hedging versus Bartlett delta and practitioner delta hedging expressed as the percentage improvement with respect to the objective function $Y(0)$.}
\label{Table:Tab4}
\end{table}

 We extended the stochastic volatility analysis by using the Black--Scholes model with a constant volatility of 20\% to value the option in the accounting P\&L approach. We find that the results are almost as good as those in Exhibits 6 and 7. This is evidence that a simple pricing model can be used in conjunction with more complex asset-price processes to develop hedging strategies.

\section{Conclusions}

This paper has explained how reinforcement learning can be used to produce an optimal hedging strategy when a particular stochastic process has been specified. We find that the accounting P\&L approach works better than the cash flow approach. Our results indicate that good results can be obtained when the derivative valuation model used in the accounting P\&L approach is simpler than the model  used to generate asset prices. This suggests that, in practice, hedging strategies can be developed by using a simple pricing model in conjunction with more complex asset-pricing processes.

The impact of trading costs is to under-hedge relative to delta hedging in some situations and over-hedge relative to delta hedging in other situations. When the current holding is lower that the holding recommended by delta hedging traders should usually under-hedge. When the reverse is true they should usually over-hedge.

A novel feature of our approach is that we build on research in the machine learning literature to provide an accurate estimate of the standard deviation of the cost of hedging. Specifically, we show how standard reinforcement learning algorithms can be extended so that both the first and second non-central moments of the probability distribution of the cost of hedging are estimated and used in the objective function. If skewness or kurtosis are a concern our approach can potentially be extended so that an objective function incorporating higher non-central moments is used. 

An extension of this research could be to allow transaction costs  to be stochastic. In normal markets transaction costs are often quite low, but a shortage of liquidity can lead to sharp increases. 
The reinforcement learning approach is also likely to be particularly useful for hedging volatility exposure as this leads to quite high transaction costs. (Volatility is usually assumed to follow a mean reverting process which is likely to have implications somewhat similar to those illustrated in Exhibit 3 for the hedging strategy.) Exotic options, which can be quite difficult to hedge using the standard Greek letter approach, may also benefit from reinforcement learning.

Our approach can be used to identify hedging strategies that work well for a range of asset-price processes even when valuation models for the processes are not known. (This is the case even in situations where there are no trading costs.) One idea is to use recent research concerned with generating synthetic data that reflects market data. Another is to use a mixture model where several stochastic processes are specified and the decision maker samples randomly to determine which process will apply before each episode is generated.

\bibliographystyle{nonumber}

\end{document}